\documentclass[aps,prl,twocolumn,showpacs,superscriptaddress,floatfix]{revtex4}
\usepackage{graphicx,ulem}

\begin{document}
  \bibliographystyle{prsty}

  \title{Size-dependent Surface States on Strained Cobalt Nanoislands on Cu(111)}
  \author{M.\ V.\ Rastei}
  \author{B.\ Heinrich}
  \author{L.\ Limot}
  \affiliation{Institut de Physique et Chimie des Mat\'{e}riaux de Strasbourg$\text{,}$ UMR 7504$\text{,}$ Universit\'{e} Louis Pasteur, F-67034 Strasbourg, France}
  \author{P.\ A.\ Ignatiev}
  \author{V.\ S.\ Stepanyuk}
  \author{P.\ Bruno}
  \affiliation{Max-Planck-Institut f\"{u}r Mikrostrukturphysik, Weinberg 2, D-06120 Halle/Saale, Germany}
  \author{J.\ P.\ Bucher}
  \affiliation{Institut de Physique et Chimie des Mat\'{e}riaux de Strasbourg$\text{,}$ UMR 7504$\text{,}$ Universit\'{e} Louis Pasteur, F-67034 Strasbourg, France}
  \date{\today}
  \begin{abstract}
Low-temperature scanning tunneling spectroscopy over Co nanoislands on Cu(111) showed that the surface states of the islands vary with their size. Occupied states exhibit a sizeable downward energy shift as the island size decreases. The position of the occupied states also significantly changes across the islands. Atomic-scale simulations and \textit{ab inito} calculations demonstrate that the driving force for the observed shift is related to size-dependent mesoscopic relaxations in the nanoislands.
  \end{abstract}
  \pacs{61.46.-w, 73.20.At, 73.22.-f}
  \maketitle
\begin{figure}[t]
\includegraphics[bbllx=84,bblly=308,bburx=465,bbury=790,width=0.44\textwidth,clip=]{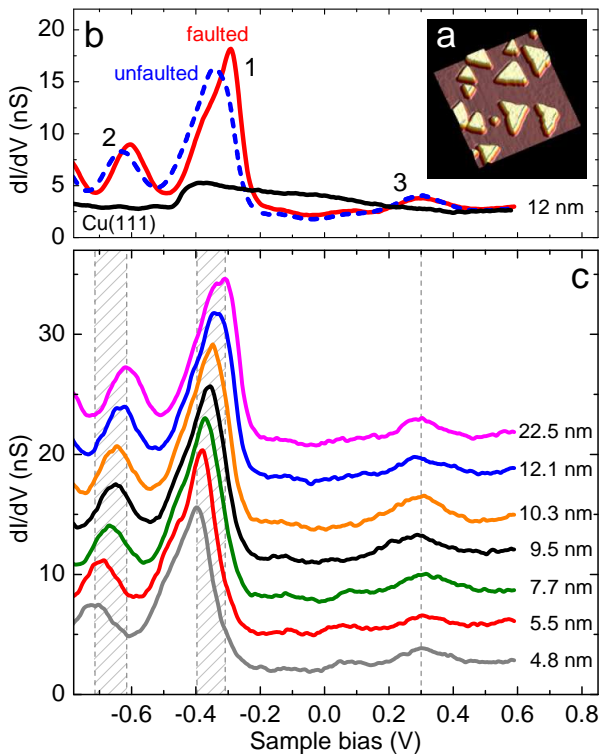}
    \caption[fig1]{(color online). a) Pseudo-three-dimensional representation of a constant-current STM image of Co nanoislands on Cu(111) ($100 \times 100$ nm$^2$, $1.5$ nA, $0.6$ V). b) $dI/dV$ spectra over Co nanoislands of opposite orientation (averages of $60$ spectra acquired on distinct $12$ nm islands) and over Cu(111). Feedback loop opened at $1.5$ nA, $0.6$ V. c) $dI/dV$ spectra on unfaulted nanoislands of increasing size. Feedback loop opened at $1.5$ nA, $0.6$ V. Spectra over islands sizes of 5.5, 7.7, 9.5, 10.3, 12.1, 22.5 nm are vertically shifted upward by 3, 6, 9.2, 12, 16, 19 nS, respectively. The hatched areas delimit the range over which an energy shift is observed for peak \textbf{1} and \textbf{2} in these spectra. The dashed line is positioned at peak \textbf{3}.}
    \label{fig1}
\end{figure}

It was recognized quite early that metallic particles exhibit unique properties that differ significantly from their bulk counterparts \cite{per81}. Nanoislands grown on metal surfaces, in particular, have been a matter of intense research for decades in view of prospective applications in a vast variety of domains ranging from magnetoelectronics, catalysis, optoelectronics, to data storage technology. The electronic, magnetic and chemical properties of a nanoisland are governed by the size, shape and structure of the island. These, in turn, are profoundly influenced by the lattice mismatch with the metal substrate and, for heteroepitaxial systems, also by the bonding interactions in the island/substrate interface (ligand effects). Metal islands tend to adopt the lattice parameter of the underlying surface \cite{ker97}, and as a consequence the bond lengths between the metal atoms in the supported nanoisland are different than those in the parent metals, resulting in changes due to strain. A theoretical study on homoepitaxial double layer Cu islands on Cu(111) \cite{lys02}, has shown that the strain produces an inhomogeneous distribution of bond lengths over the nanoisland, the average bond length varying with island size. The nanoislands also locally distort the surface and induce a displacement pattern in the substrate which affects the diffusion of atoms and, ultimately, the growth of the nanoislands.\\
\- Despite these studies, our knowledge of how strain affects the properties of a metallic nanoisland, especially the electronic states, remains very limited. The fundamental problem of the change in energy upon lattice distortion in solids has first been addressed by J. Friedel within a simple model \cite{fri69}, and latter on extended to various problems of lattice contractions at metal surfaces and clusters \cite{des93}. On the same lines, it has been shown for thin films that strain effects, along with ligand effects, can cause a shift of the surface $d$ band \cite{mav98,kit04,cal05}, resulting in chemical properties that are significantly different from those of the pure overlayer metal. Recently, a modification of electronic states due to a local strain field induced by a nanopattern formation has been observed for Cu(100) covered with N atoms \cite{sek07}.\\
\- In this Letter, we specifically focus on the interplay between strain-induced structural relaxations and the surface states of Co nanoislands on Cu(111). These nanoislands constitute a reference system that has been extensively investigated by Scanning Tunneling Microscopy/Spectroscopy (STM/STS) \cite{vaz00,die03,pie04,yay07,pie06}. By acquiring STS data over islands of increasing size, we establish that the occupied surface states exhibit a size-dependent energy shift, $i.\ e.$ that the $d$-like band shifts in energy. A shift is also observed at the corners and edges of the island with respect to the center of the island. Atomic-scale simulations and \textit{ab initio} calculations demonstrate that the energy positions of the occupied states are determined by mesoscopic relaxations in the nanoislands. Our work suggests that surface states can be a sensitive probe for variations of the atomic structure at the nanoscale.\\  
\- The measurements were performed in a modified Omicron ultrahigh vacuum STM ($<10^{-10}$ mbar) cooled to 4.6 K. The single-crystal Cu(111) substrate was cleaned by repeated cycles of Ar$^+$ sputtering and anneal to $500^{\circ}$C. About $0.7$ monolayers (ML) of Co were evaporated at $0.15$ ML min$^{-1}$ on to the Cu(111) surface at room temperature from a thoroughly outgassed Co rod. After deposition, the sample was immediately transferred in the pre-cooled STM. Triangular-like Co nanoislands two atomic layers high are then observed (Fig.~\ref{fig1}a), with opposite orientations in a ratio of $3:2$. Following \cite{vaz00}, the majority population have a fcc stacking as Cu(111) (unfaulted islands), whereas the minority population contain a stacking fault (faulted islands). The spectra of the differential conductance of the tunneling junction, $dI/dV(V)$, where $V$ is the sample bias measured with respect to the tip, were acquired via lock-in detection with a bias modulation of $3$ mV rms at $\approx 5$ kHz (feedback loop open). A variety of etched W tip were employed. After a sputter/anneal cycle, the tips were treated by soft indentations into the clean copper surface, until tip-structure artifacts were minimized in the $dI/dV$ spectra over the voltage range of interest. The step-like onset of the Cu(111) Shockley surface state appeared then as a sharp and reproducible feature in the $dI/dV$ spectra (Fig.~\ref{fig1}b).
Typical spectra acquired in the center of faulted and unfaulted islands with sizes of $12$ nm are presented on Fig.~\ref{fig1}b. The size of the island is measured in terms of the length of its edge, which, for conveniency, is determined by evaluating the area of the island and assuming that the island is an equilateral triangle. Three distinct peaks are observed, two below the Fermi energy and one above in agreement with \cite{pie04}. The occupied states on unfaulted islands present a dominant peak at $-0.31$ V (labeled \textbf{1}) which has its origin in the minority band of Co islands \cite{die03,pie04}. The same peak falls at a higher energy of $-0.28$ V on faulted islands. Given the asymmetric line shape, which is moreover tip-dependent, for consistency the peak positions were evaluated taking the center of gravity of the line. At lower energies, a second peak (labeled \textbf{2}) is present (unfaulted: $-0.64$ V, faulted: $-0.60$ V), whose amplitude is strongly tip-dependent. Finally, the unoccupied states on both type of islands present a peak at $+0.3$ V (labeled \textbf{3}).\\
\begin{figure}[t]
\includegraphics[bbllx=58,bblly=488,bburx=454,bbury=733,width=0.44\textwidth,clip=]{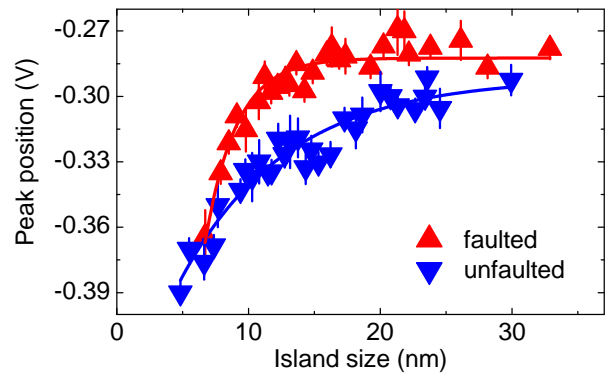}
    \caption[fig2]{(color online). Peak position (labeled \textbf{1}) versus island size (up triangles: faulted, down triangles: unfaulted). Solid lines are a guide to the eye. Data is binned by steps of $0.5$ nm.}
    \label{fig2}
\end{figure}
Figure~\ref{fig1}c illustrates the main experimental finding. As shown here for unfaulted islands, peak \textbf{1} and \textbf{2} shift downward in energy when the island size decreases from $22.5$ nm to $4.8$ nm, with no appreciable changes in the shape and in the amplitude of the line. On the contrary, the position of \textbf{3} is unchanged. A similar behavior is observed over faulted islands. Figure~\ref{fig2} presents a quantitative evaluation of the size-dependent shift of peak \textbf{1}. Spectra were acquired in the center of $230$ islands of increasing island size -- from $4.8$ to $31.9$ nm on unfaulted islands, and from $6.7$ to $32.9$ nm on faulted ones. A monotonical shift over $0.09$ V is observed for unfaulted islands, whereas the faulted ones show a steeper increase of the shift with an asymptotic behavior appearing above island larger than $13$ nm.\\
\- In order to explore a possible involvement of edge effects, especially in the smallest islands studied, spectra were acquired over the surface of faulted and unfaulted islands. Figure~\ref{fig3}a shows the typical spatial dependency of peak \textbf{1} when moving from a corner of an island (here a faulted one), through the center, to the opposite edge. The peak positions extracted from all the acquired spectra, along with the line profile of the island, are presented in Figs.~\ref{fig3}b and~\ref{fig3}c, respectively. In the center of the island (spectra $5$, $6$, $7$) peak \textbf{1} is shifted approximately to $-0.36$ V because of the small island size ($7.1$ nm). However, at $\approx 1.0$ nm from the edge peak \textbf{1} starts to move further downward in energy (spectrum $8$ on Fig.~\ref{fig3}a), reaching a displacement of $-0.03$ V (spectrum $9$) relative to the center position. A diminished amplitude is also observed, in agreement with \cite{pie06}. Similarly, at $\approx 2.5$ nm from the corner, an additional displacement of $-0.05$ V progressively sets in (spectra $2$, $3$, $4$), the amplitude of the peak also decreasing until disappearance at the corner (spectrum 1). This additional shift only occurs close to the edges and corners, and has a negligible influence on the shift in the center of the island, even in the smallest islands investigated. In a previous work it was concluded that Co-Co bond lengths must vary with the lattice constant of a given substrate, affecting thereby the energy positions of the occupied peaks \cite{ras07}. This strongly hints to a size-dependent in-plane Co-Co bond variation for Co nanoislands on Cu(111), $i.\ e.$ to a mesoscopic relaxation, as we establish below.\\
\begin{figure}[t]
\includegraphics[bbllx=44,bblly=390,bburx=573,bbury=667,width=0.46\textwidth,clip=]{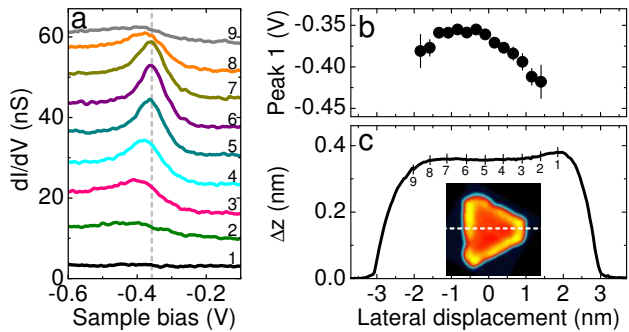}
    \caption[fig3]{(color online). a) $dI/dV$ spectra acquired across a faulted $7.1$ nm nanoisland. Feedback loop opened at $1$ nA, $0.5$ V. Spectra $2$--$9$ are shifted vertically by steps of $7$ nS. The dashed line is centered on the peak positions of spectra acquired in the island center of gravity (spectra $5$--$7$).  b)  Energy shift of peak \textbf{1}, and c) Line profile across the island (from corner to edge as depicted by the dashed line on the image). Numbers from 1 to 9 are the corresponding positions were the spectra noted 1 to 9 on panel a) were acquired.}
    \label{fig3}
\end{figure}
A set of calculations were conducted firstly to find the fully relaxed atomic configurations, and secondly to estimate the corresponding energy shift of peak \textbf{1}. The relaxed atomic configurations have been obtained by means of \textit{ab initio} fitted many-body potentials formulated in the second moment approximation of the tight-binding theory \cite{lys02,ros89,ste00}. Perfectly triangular unfaulted bilayer Co islands with sizes ranging from $4$ nm up to $30$ nm were studied. The distribution of the topmost Co layer in-plane bond length over a $15$ nm island is presented in Fig.~\ref{fig4}a as an example. The distribution is inhomogeneous over the island. The Co atoms at the edges/corners are relaxing in the direction of the center of the island and take other equilibrium positions with shorter bonds with respect to Co atoms in the center. The inner region around the gravity center of the island presents a nearly homogeneous distribution of the bond lengths (inset of Fig.~\ref{fig4}a), and thus an average in plane bond length $r$ and interlayer distance $z$ can be used to describe the structure of this region. Both $r$ (Fig.~\ref{fig4}b) and $z$ (not shown) depend on the island size. With increasing island size, $r$ increases following a close-to-exponential behavior towards the ideal bond length of bulk Cu ($r_0=0.2556$ nm). The mismatch $(r_0-r)/r_0$ varies from $0.1\%$ and approaches in the smallest islands the macroscopic mismatch of $2\%$. Concomitantly, the interlayer distance decreases about $1\%$ with increasing island size.\\
\begin{figure}[t]
\includegraphics[bbllx=85,bblly=260,bburx=493,bbury=692,width=0.44\textwidth,clip=]{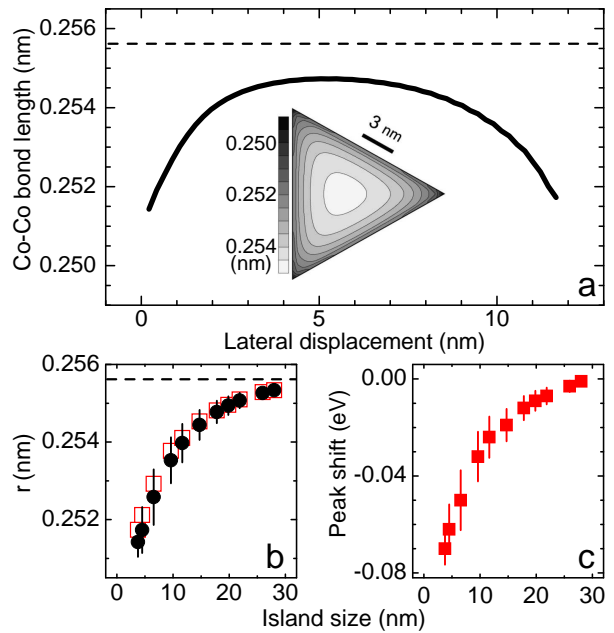}
    \caption[fig4]{(color online). a) Variation of the in-plane bond length of the top Co layer across a $15$ nm island (dashed line: ideal bulk Cu bond). Inset: Spatial distribution of in-plane Co-Co bonds. b) Average-bond length $r$, described in the text, for the top (solid circles) and bottom (open squares) Co layers, and, c) Energy shift of peak \textbf{1} in the center of islands with increasing size.}
    \label{fig4}
\end{figure}
In order to model the experimental local density of states (LDOS) above the Co islands and link the Co-Co bond relaxations to the electronic structure change, the Korringa-Kohn-Rostoker Green's function method based on density functional theory was used as in previous studies \cite{die03,ste95,zel95,zab05}. The method was improved by employing the full potential approximation. The surface states above the center of each island were estimated by performing calculations for an infinite Co bilayer. Spectral density maps (imaginary part of momentum-resolved energy-dependent Green's function) were then plotted to study in detail the surface-states band structure and gain some insight on the peaks observed. Peak \textbf{1} and \textbf{2} have their origin in the same minority $d_{3z^2-r^2}$ band splitted at the intersection with the Cu(111) bulk band. The hybridization of $s-p$ states with the minority $d_{3z^2-r^2}$ band situated in the bulk band gap form peak \textbf{1}. Peak \textbf{2} appears due to a resonant overlapping of Cu(111) bulk states with the Co minority $d_{3z^2-r^2}$ band. Peak \textbf{3} is derived from the unoccupied Co minority $d$ band. All peaks are located in the first half of the Brillouin zone.\\ 
\- To account for the size-dependent Co-Co bond relaxation in the LDOS, the in-plane lattice constant of the infinite Co bilayer was fixed to $r$, and the interlayer distance to $z$. The calculated shift of peak \textbf{1} versus the island size is presented in Fig.~\ref{fig4}c and is seen to vary over $0.07$ eV, in good agreement with the experimental results. Peak \textbf{2} yields a similar variation, also in agreement with the experimental results. The simultaneous shift of both peaks, which are located in distinct points of the occupied $d_{3z^2-r^2}$ band, indicates that strain effects are causing an energy shift of the band. The variation of the in-plane bond length is the driving force for the observed shift. This finding agrees with recent STS data acquired over Co nanoislands on Pt(111) \cite{mei06}, and on Au(111) \cite{ras07}, where peaks of same origin as peak \textbf{1} are observed. An increased mismatch of $9.4\%$ (Pt) and of $13\%$ (Au), $i.\ e.$ a linear increase of the lateral expansion of the nanoislands, is sufficient to explain the displacement of these peaks relative to Co/Cu(111), at positions of approximately $-0.23$ V and $-0.15$ V, respectively. This also indicates that the ligand effect contributes little to the shift in these systems.\\ 
\- The downward shift of peaks \textbf{1} and \textbf{2} for decreasing island size can be rationalized in the framework of a tight-binding model. Taking an infinite Co bilayer, the shift of the band as a function of the Co-Co bond length is given by $\Delta E(\overrightarrow{k},r)=\beta(r)\: F(\overrightarrow{k},r)$, where $F(\overrightarrow{k},r)$ is a positive sum of k-dependent cosine functions ($\|\overrightarrow{k}\|<\pi/2r$). Following \cite{fri69}, we express the transfer integral as $\beta=\beta_0\: exp(-q\:r)$, where $q$ is a positive material-dependent constant, and neglect the crystal-field contribution. For occupied electronic states the transfer integral is negative ($\beta_0<0$), and in the limit of a small variation of the average bond length ($r=r_0-\delta r$) as in Fig.~\ref{fig4}b, it follows that $\Delta E \propto (1+q\:\delta r)\:\beta_0$. Peak \textbf{1} and \textbf{2} both exhibit a negative shift close in amplitude, which varies linearly with the Co-Co bond length (Figs.~\ref{fig4}b and ~\ref{fig4}c). On the contrary, the unoccupied minority $d$ band (peak \textbf{3}) is likely not shifting in energy. This follows from the flat nature of the unoccupied minority band around the $\Gamma$-point, which then implies a narrow bandwidth ($\beta_0$ is close to zero) and, hence, a negligible energy shift.\\
\- Finally, calculations were also performed to reproduce the experimental shift at the edges/corners of the island. Despite the lower accuracy expected given the spatial averaging method employed, a shift of $-0.03$ eV and $-0.06$ eV is estimated for peak \textbf{1} at the edges and corners, respectively, in agreement with the experimental values of Fig.~\ref{fig3}b.\\
\- In summary, the impact of atomic relaxations in Co nanoislands on the energy position of the minority surface state were revealed. Although evidenced for a particular system, this result confirms the predictions on mesoscopic relaxation \cite{lys02,ste00}. Our results give a clear evidence that the surface-states electrons on nanoislands are significantly affected by local atomic structure. When the size of the island decreases, $i.\ e.$ with increasing lateral strain, the occupied states move to lower energies. A variation of the catalytic activity can then be expected with island size \cite{mav98}.\\  
\- The Strasbourg authors thank the NoE MAGMANet for financial support. This work has been supported by the Deutsche Forschungsgemeinschaft SPP 1153 and SPP 1165.\\

\end{document}